\documentclass[%
reprint,
superscriptaddress,
 amsmath,amssymb,
pra,
]{revtex4-2}

\usepackage{graphicx}
\usepackage{dcolumn}
\usepackage{bm}
\usepackage{color}
\usepackage{comment}

\usepackage{bbm}
\usepackage{mathtools}
\usepackage{siunitx}
\usepackage{physics}
\usepackage{enumerate}

\usepackage[colorlinks=true,bookmarks=false,allcolors=blue]{hyperref}
\hypersetup{pdfstartview={FitH},pdfpagemode={UseNone}, breaklinks=true,
            colorlinks,allcolors=blue, bookmarksopen=true, pdfnewwindow=true}
\usepackage[all]{hypcap}

\begin{document}

\preprint{APS/123-QED}

\title{Fundamental resolution limit of quantum imaging with undetected photons}

\author{Andres Vega}%
 \email{andres.vega@uni-jena.de}
 \affiliation{
Institute of Applied Physics, Abbe Center of Photonics, Friedrich Schiller University Jena, Albert-Einstein-Str. 15, 07745 Jena, Germany}%
\author{Elkin A. Santos}%
 \email{elkin.santos@uni-jena.de}
 \affiliation{
Institute of Applied Physics, Abbe Center of Photonics, Friedrich Schiller University Jena, Albert-Einstein-Str. 15, 07745 Jena, Germany}%
\author{Jorge Fuenzalida}
\affiliation{
Fraunhofer Institute for Applied Optics and Precision Engineering IOF, Albert-Einstein-Str. 7, 07745 Jena, Germany}
\author{Marta Gilaberte~Basset}
\affiliation{
Fraunhofer Institute for Applied Optics and Precision Engineering IOF, Albert-Einstein-Str. 7, 07745 Jena, Germany}
\author{Thomas Pertsch}
\affiliation{
Institute of Applied Physics, Abbe Center of Photonics, Friedrich Schiller University Jena, Albert-Einstein-Str. 15, 07745 Jena, Germany}%
\affiliation{
Fraunhofer Institute for Applied Optics and Precision Engineering IOF, Albert-Einstein-Str. 7, 07745 Jena, Germany}
\author{Markus Gräfe}
\affiliation{
Fraunhofer Institute for Applied Optics and Precision Engineering IOF, Albert-Einstein-Str. 7, 07745 Jena, Germany}
\author{Sina Saravi}
\affiliation{
Institute of Applied Physics, Abbe Center of Photonics, Friedrich Schiller University Jena, Albert-Einstein-Str. 15, 07745 Jena, Germany}%
\author{Frank Setzpfandt}
\affiliation{
Institute of Applied Physics, Abbe Center of Photonics, Friedrich Schiller University Jena, Albert-Einstein-Str. 15, 07745 Jena, Germany}%
\affiliation{
Fraunhofer Institute for Applied Optics and Precision Engineering IOF, Albert-Einstein-Str. 7, 07745 Jena, Germany}

\date{\today}

\renewcommand{\d}{{\rm d}}
\renewcommand{\P}{{\rm P}}
\renewcommand{\S}{{\rm S}}
\newcommand{\I}{{\rm I}}
\renewcommand{\A}{{\rm A}}
\newcommand{\B}{{\rm B}}

\begin{abstract}
Quantum imaging with undetected photons relies on the principle of induced coherence without induced emission and uses two sources of photon-pairs with a signal- and an idler photon.
Each pair shares strong quantum correlations in both position and momentum, which allows to image an object illuminated with idler photons by just measuring signal photons that never interact with the object. In this work, we theoretically investigate the transverse resolution of this non-local imaging scheme through a general formalism that treats propagating photons beyond the commonly used paraxial approximation. We hereby prove that the resolution of quantum imaging with undetected photons is fundamentally diffraction limited to the longer wavelength of the signal and idler pairs.
Moreover, we conclude that this result is also valid for other non-local two-photon imaging schemes.
\end{abstract}

\maketitle
\section{Introduction}

Entangled photons can be generated by spontaneous parametric down-conversion (SPDC) \cite{HONGMANDEL.TheorySPDC.1985}, where a pump photon impinges onto a second-order nonlinear crystal and is converted into a signal and an idler photon that share quantum correlations simultaneously in, for example, transverse momentum and position. This quantum correlation enabled two unconventional imaging schemes \cite{Gilaberte.Review.2019}, quantum ghost imaging \cite{SHIH.QGhostImag.1995, SHIH.QGhostDiff.1995} and quantum imaging with undetected photons (QIUP) \cite{LEMOS.QIUP.2014,LAHIRI.TheoryQIUP.2015, Kviatkovsky.MicroscopyQIUPMIR.2020,GILABERTE.VideoQIUP.2021,Toepfer.2022, Paterova.Hyperspectral.2020}, that produce a so-called ``non-local'' image by letting only the idler photon interact with the object while the camera measures the non-interacting signal photon. The most notable feature of these non-local imaging schemes is, that they enable the use of two-color photon-pairs, i.e., signal and idler with non-degenerate wavelengths. This unique characteristic allows to overcome complications of the detection in certain wavelength ranges where sensors have low efficiency \cite{BOYDASPDEN.PhotonSparse.2015,SHIH.TwoColorResolvingPower.2010,Boyd.TwoColorGhost.2009}, and has tremendous potential for bio-sensing, where sensitive samples can be imaged using conventional single photon cameras in the visible range, while the sample is being illuminated by photons with much lower energy. 

It is a well known fact that the resolution of a diffraction-limited classical imaging scheme depends on the wavelength $\lambda$ of the particle that interacts with the object \cite{Abbe.resolution.1873,RAYLEIGH.Resolution.1879,born_wolf}. Assuming that the optical elements have a numerical aperture NA$=1$, the resolution is $\approx \lambda/2$.
It is of great interest to investigate two-color quantum imaging configurations to determine the exact role of both wavelengths in the resolution. Although several works have discussed the resolution of quantum imaging \cite{DANGELO.ResolutionGhost.2005,Boyd.TwoColorGhost.2009,Moreau.ResolutionGhost.2018,Moreau.ResolutionGhostPopper.2018, FUENZALIDA.resolutionQIUP.2020,Kviatkovsky.MicroscopyQIUPMIR.2020,LAHIRI.QIUPosRes.2021, RAMELOW.QIUPPossExp.2021}, they have treated this question only within the paraxial regime. This fits very well with most experiments relying on commercially available nonlinear crystals. Their typical thickness is much larger than the wavelengths of signal and/or idler, which allows the modes of signal and idler to cover only a small range of transverse momenta. However, the recent advent of thinner nonlinear materials as photon-pair sources \cite{CHEKHOVA.NonlinearFilms.2021, CHEKHOVA.SPDCMetasurface.2021, CHEKHOVA.WithoutMomCons.2019} opens up the possibility to have photons in a momentum range beyond the paraxial regime as the crystal thickness can be smaller than the wavelengths of the down-converted photons. This pushes the need to have a more general description of the resolution suitable in the non-paraxial regime.

\begin{figure*}[ht]
\centering\includegraphics{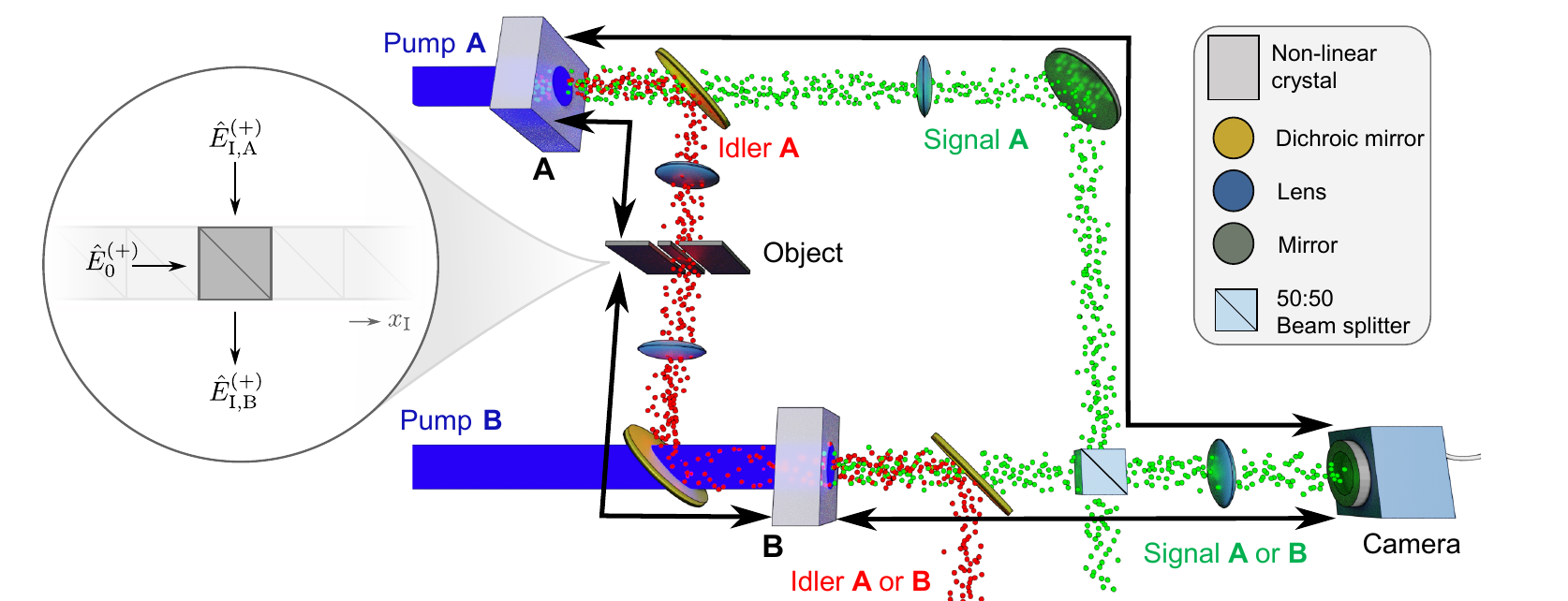}
\caption{Sketch of the setup of quantum imaging with undetected photons (QIUP) based on position correlations. In the idler arm, the central plane of crystal A is imaged onto the object by a system (e.g., a 4f configuration) with magnification one represented by the arrow and also the object is imaged onto the central plane of crystal B. In the signal arm, the central planes of both crystals are imaged onto the camera. The inset depicts the model of the object that consist of a beam splitter at each $x_\I$ position.}
\label{fig:setup}
\end{figure*}

In this work, we derive a general analytical model for describing QIUP, that goes beyond the paraxial regime and allows us to theoretically identify the diffraction-limited resolution of this imaging scheme. To this end, we first introduce the formalism to derive the two-photon state in a non-paraxial framework. Afterwards, we  use the state to study the response of QIUP to two slits of infinitesimal width and find the minimum resolvable distance between them. This work is focused on the scheme enabled by position correlations of signal and idler photons \cite{LAHIRI.QIUPPosition.2021}, as depicted in Fig.~\ref{fig:setup}. In the idler arm, the central plane of source~A is imaged on the object by an imaging system with magnification equal to one represented by an arrow. In the same manner, the object plane is imaged onto the central plane of source~B. In the signal arm, the central planes of both sources are imaged onto the camera.

The manuscript is organized as follows. In Sec.~\ref{sec:twophoton} we introduce a non-paraxial description of the two-photon state generated through SPDC. In the next Sec.~\ref{sec:QIUP}, we derive the general expression of the non-local image. In Sec.~\ref{sec:resolution}, a numerical and an analytical model for the resolution are analyzed. Lastly, in Sec.~\ref{sec:disc} we present an additional discussion on the diffraction-limited resolution of other non-local two-photon imaging configurations.

\section{Two-photon state beyond the paraxial regime}
\label{sec:twophoton}

\subsection{Momentum Representation}
To construct a rigorous general framework, we consider the positive-frequency part of the electric field operator in free space, which in the interaction picture reads as \cite{TSANG.QuantLitho.2007}
\begin{equation}
\begin{split}
    \hat{\mathbf{E}}^{(+)} (\mathbf{r},t) = \frac{i}{(2\pi)^{3/2}} & \sum_s \int \dd\mathbf{k} \left( \frac{\hbar \omega}{2 \varepsilon_0} \right)^{1/2} \\& \times \hat{a}(\mathbf{k},s) \mathbf{e}(\mathbf{k},s) e^{i (\mathbf{k}\cdot\mathbf{r} - \omega t)}, \label{eq:fullEop}
\end{split}
\end{equation}
where the wave-vector integral is $\int \dd\mathbf{k}= \int_{-\infty}^{+\infty} \dd k_z \int_{-\infty}^{+\infty} \dd k_y \int_{-\infty}^{+\infty} \dd k_x$, $s \in \{1,2\}$ refers to the two possible polarizations perpendicular to each wave-vector $\mathbf{k}$, $\hat{a}$ is the annihilation operator, $\mathbf{e}$ is the unit-vector along each of the two polarization directions, $\omega=c |\mathbf{k}|=c\sqrt{k_x^2+k_y^2+k_z^2}$ is the angular frequency, $\epsilon_0$ is the vacuum permittivity, and $c$ is the speed of light in vacuum. Moreover, the annihilation and creation operators obey the commutation relations
\begin{equation}
    [\hat{a}(\mathbf{k},s), \hat{a}^\dagger(\mathbf{k}',s')] = \delta (\mathbf{k} - \mathbf{k}') \delta_{ss'} .
\end{equation}
Since conservation of energy in the pair-generation process eventually forces a strict relation between the frequencies of the generated photons, we need to make frequency dependencies explicit in our description. To do this, we express the variable $k_z$ in terms of $\omega$.  Additionally, only positive $k_z$ are relevant in this setup, resulting in $k_z=+ \sqrt{(\omega/c)^2-k_x^2-k_y^2}$. The corresponding integral is then limited to $\int_{0}^{+\infty} \dd k_z$. Hence, $\dd k_z$ can be written as 
\begin{equation}
    \dd k_z = \dd \left[ \left( \frac{\omega}{c} \right)^2-k_x^2-k_y^2 \right]^{1/2} = \frac{\omega}{c^2 k_z} \dd \omega\,,
\end{equation}
and the integral follows the form
\begin{equation}
    \int \dd \mathbf{k} \longrightarrow \int_{0}^{+\infty} \dd \omega \int_{-\omega/c}^{+\omega/c} \dd k_x \int_{-\sqrt{\frac{\omega^2}{c^2}-k_x^2}}^{+\sqrt{\frac{\omega^2}{c^2}-k_x^2}} \dd k_y \, \frac{\omega}{c^2 k_z}.
\end{equation}
Noteworthy, the $k_x, k_y$ integrals are restricted to the region $k_x^2+k_y^2 \leq \omega^2/c^2$ of propagating plane-waves. Hence, with a fixed value of $k_x$, the $k_y$ integral has a range limited to $\pm \sqrt{(\omega/c)^2-k_x^2}$.
Furthermore, since the commutation relations change under the change of variable, 
\begin{equation}
    \begin{split}
         [\hat{a} & (\mathbf{k},s), \hat{a}^\dagger(\mathbf{k}',s')]  \\& = \delta (k_x - k_x') \delta (k_y - k_y') \delta (k_z - k_z') \delta_{ss'} \\& = \delta (k_x - k_x') \delta (k_y - k_y') \delta (\omega - \omega') \delta_{ss'} \frac{c^2 k_z}{\omega},
    \end{split}
\end{equation}
the annihilation operator can be newly defined as
\begin{equation}
     \hat{a}(k_x,k_y,\omega,s) = \left(\frac{\omega}{c^2 k_z}\right)^{1/2} \hat{a}(k_x,k_y,k_z,s),
\end{equation}
which results in the commutation relations
\begin{equation}
\begin{split}
    [\hat{a} & (k_x,k_y,\omega,s), \hat{a}^\dagger(k_x',k_y',\omega',s')]  \\ & = \delta (k_x - k_x') \delta (k_y - k_y') \delta (\omega - \omega') \delta_{ss'} .
\end{split}
\end{equation}
To better understand the physics and also to reduce the weight of the numerical calculations in the non-paraxial regime of the signal and idler correlations in the next sections of this work, we resort to only one transverse dimension $k_x$ without compromising the underlying physics. Such simplification corresponds to only considering photons with $k_y\approx 0$, which is the least limiting case for the range of $k_x$, therefore, the scenario with $k_y\approx 0$ is suitable for finding the best possible transverse resolution. The variable $k_x$, from here on denoted as $q$, represents then the transverse momentum and $x$ is its corresponding transverse position dimension.
As mentioned in Ref.~\cite{TSANG.QuantLitho.2007}, to do this reduction in dimension formally, we use the substitution $\int \dd k_y\rightarrow 2\pi/l_y$, where $l_y$ is a normalization length scale in the $y$-direction. Consequently, we redefine the annihilation operator as $\hat{a}(k_x,k_y,\omega,s)\rightarrow \hat{a}(q,\omega,s)\sqrt{l_y/2\pi}$ which satisfies $[\hat{a}(q,\omega,s),\hat{a}^\dagger(q',\omega',s')]=\delta(q-q')\delta(\omega-\omega')\delta_{ss'}$.

Combining all the terms, the electric field operator becomes $\hat{\mathbf{E}}^{(+)} (\mathbf{r},t) = \int_{0}^{+\infty} \dd \omega \:
    \hat{\mathbf{E}}^{(+)} (\mathbf{r},\omega) \exp(- i \omega t) $, which under the aforementioned assumption takes the form
\begin{equation}
    \begin{split}
        \hat{\mathbf{E}}^{(+)} (x,z,\omega) = & i \left[\left( \frac{\hbar}{16 c^2 \pi^3 \varepsilon_0} \right) \left(\frac{2\pi}{l_y} \right) \right]^{1/2} 
        \\ & \times
         \sum_s \int_{-\omega/c}^{+\omega/c} \dd q \Big[ \frac{\omega}{k_z^{1/2}} \hat{a}(q,\omega,s) 
         \\ &  \quad \quad \times 
          \mathbf{e}(q,\omega, s) \exp \left(i qx + i k_z z \right) \Big]
          \,.
    \label{eq:Eoperator}
    \end{split}
\end{equation}

Notice that in the paraxial regime, where $q^2\ll\omega^2/c^2$, the term $1/k_z^{1/2}$ in Eq.~(\ref{eq:Eoperator}), with $k_z(q)=\sqrt{(\omega/c)^2-q^2}$, can be taken to be approximately independent of $q$ \cite{TSANG.QuantLitho.2007}. However, to fully investigate the limit in resolution, we need to consider the full range of propagating spatial frequencies that are involved in the imaging system, including $q^2\approx\omega^2/c^2$. To study this non-paraxial region, we show in our work that it is important to carefully handle the term $k_z^{-1/2}$ that tends to infinity as the transverse momentum $q$ increases. 

Now, we calculate the two-photon state in the low-gain regime. The state, removing its vacuum component, has the general form 
\begin{equation}\label{eq:SPDCstate}
        \begin{split}
        \ket{\psi} & \propto 
        \int \dd t \int \dd \mathbf{r}  
        \sum_{\alpha,\beta,\gamma} \Big[ \chi^{(2)}_{\alpha\beta\gamma} (\mathbf{r}) 
        \\ & \times 
        E_{\P,\gamma} (\mathbf{r},t) \hat{E}^{(-)}_\alpha (\mathbf{r},t) \hat{E}^{(-)}_\beta (\mathbf{r},t) \ket{0,0} \Big],
        \end{split}
\end{equation}
where $\chi^{(2)}$ is the second-order nonlinear susceptibility of the crystal, the time integral $\int \dd t$ considers the interaction time and the spatial integral $\int \dd \mathbf{r}$ covers the volume of the nonlinear crystal where the interaction occurs. The coefficients $\gamma, \alpha, \beta $ refer to the polarization direction of the pump, signal and idler fields, respectively. Additionally, we take the positive part of the pump in the undepleted pump approximation, treating it as a classical field defined as $\mathbf{E}_\P (\mathbf{r},t) = \int_0^{+\infty} \dd \omega_\P \mathbf{E}_\P(\mathbf{r},\omega_\P) \exp(-i \omega_\P t)$, where the transverse spatial components can be expanded into plane waves $\mathbf{E}_\P (x, z) = \int_{-\infty}^{+\infty} \dd q_\P \, \mathbf{E}_\P(q_\P; z=0) \exp(i  q_\P x + i k_{z \P} z)$. Here, we are similarly considering only modes that propagate in the positive $z$-direction and in the $x$-$z$ plane. Hence, introducing the pump field and the electric field operator of Eq.~(\ref{eq:Eoperator}) for the signal (primed) and idler (double-primed) photons into the two-photon state results in
\begin{widetext}
\begin{equation}
    \begin{split}
    \ket{\psi} \propto  &
    \int_{-\infty}^{+\infty} \dd t \int_{-\infty}^{+\infty} \dd x \int_{-L/2}^{+L/2} \dd z \sum_{\alpha,\beta,\gamma} \chi^{(2)}_{\alpha\beta\gamma} (x,z)  
    \int_{0}^{+\infty} \dd \omega_\P  \int_{-\infty}^{+\infty} \dd q_\P  E_{\P,\gamma}(q_\P,\omega_\P) \exp\left[-i (\omega_\P t - q_\P x - k_{z \P} z)\right]
     \\ & \times
    \int_{0}^{+\infty} \dd \omega' \sum_{s'} \int_{-{\omega'/c}}^{+{\omega'/c}} \dd q'  \frac{\omega'}{k_z'^{1/2}} \hat{a}^\dagger(q',\omega',s') e_\alpha(q',\omega',s') \exp \left[i(\omega' t - q'x - k'_z z) \right]
     \\  & \times
    \int_{0}^{+\infty} \dd \omega'' \sum_{s''}  \int_{-{\omega''/c}}^{+{\omega''/c}} \dd q''  \frac{\omega''}{k_z''^{1/2}} \hat{a}^\dagger(q'',\omega'',s'') e_\beta(q'',\omega'',s'') \exp \left[i(\omega'' t - q''x - k''_z z)\right]
    \ket{0,0}.
    \end{split}
    \label{General_biphoton_state}
\end{equation}
\end{widetext}
The spatial and time integrals lead to
\begin{equation}
\begin{split}
    \int_{-\infty}^{+\infty} \dd x \exp\left[i \left( q_\P - q' - q'' \right) x \right] & \propto \delta(q_\P - q' - q'') 
    \\ 
    \int_{-L/2}^{+L/2} \dd z\; 
    \exp\left[i \left( k_{z \P} - k'_{z} - k''_{z} \right) z \right] 
     & \propto  \mathrm{sinc}\left(\frac{\Delta k_z L}{2}\right), 
     \\
     \int_{-\infty}^{+\infty} \dd t \; \exp\left[-i(\omega_\P  + \omega'  + \omega'')t\right] & \propto \delta(\omega_\P-\omega'-\omega''),
\end{split}
\end{equation}
where $\Delta k_z=k_{z\P}-k'_{z}-k''_{z}$, and  $k_{z}=[(2\pi/\lambda)^2-{q}^2]^{1/2}$,  in which the pump, signal and idler have wavelengths $\lambda_\P, \lambda', \lambda''$, respectively, and $L$ is the thickness of the crystal in the $z$-direction. The crystal is assumed to be transversally much larger than the pump beam extent.
It should be emphasized that throughout this work, we take the dispersion relation of waves in free space for describing the interaction of waves in the nonlinear crystal. This is firstly to avoid focusing our calculation on a specific nonlinear crystal, and also to not having to deal with multiple reflections that can happen at the end facets between the crystal and the outside free space. Importantly, this type of treatment does not affect the main physics in our problem that is dependent on the range of generated transverse wave-vectors, since only those generated waves can escape the crystal which have a transverse wave-vector smaller than the wave-vector of the outside free space. In a realistic crystal, the main change will be in the shape of the phase-matching function, which will no longer be an exact sinc-function, as derived above, but could be a modulated function due to the Fabry-Perot effect. Nevertheless, this will not change the range of available transverse wave-vectors in the process, especially in the case of a very thin crystal, on which the main result of our work is based.

Since the pump beam is taken as a continuous-wave laser, the interaction time is taken to be infinite. By looking only at waves along $k_y\approx0$, we have the choice to fix one of the two polarization directions, say $s=1$, to always be along the $y$-direction and the other $s=2$ to be orthogonal to it. Henceforth, by properly choosing only the nonlinear component $\chi_{yyy}$ as the dominant one, all generated signal and idler along with the pump beam are only $y$-polarized, which allows us to use a scalar formulation and drop the sums over $s'$ and $s''$. 
We perform our calculations with a fixed frequency of $\omega_\S$ for the signal photons.
Taking the pump to have a very narrow spectral bandwidth around the frequency $\omega_\P$ results in idler photons at $\omega_\I=\omega_\P-\omega_\S$ following conservation of energy.
This allows us to remove the $\int_{0}^{+\infty} \dd \omega'' \int_{0}^{+\infty} \dd \omega'$ integrals over the signal and idler frequencies in Eq.~(\ref{General_biphoton_state}).
In an experimental setting, this is equivalent to a spectrally narrowband detection of signal photons, which can be achieved by placing a narrow bandpass filter of central frequency $\omega_\S$ before the camera.
Thus, the final two-photon state simplifies to 
\begin{equation}
\begin{split}
    \ket{\psi}\propto \iint_{-\infty}^{+\infty} \dd q_\S & \dd q_\I \; \Big[ \phi (q_\S,\omega_\S;q_\I,\omega_\I) 
    \\ & \times \hat{a}^\dagger(q_\S,\omega_\S) \hat{a}^\dagger(q_\I,\omega_\I)\ket{0,0} \Big],
    \label{eq:SPDCstate_q}
\end{split}
\end{equation}
with the joint transverse momentum amplitude
\begin{equation}
\begin{split}
        \phi  (q_\S,q_\I) & =  \; E_\P(q_\S + q_\I) \, \mathrm{sinc}\left( \Delta k_z \frac{L}{2}\right)
        \\ & \times 
        \left[k_{z\S}(q_\S) k_{z\I}(q_\I) \right]^{-1/2} \\ & \times 
        \mathrm{rect} \left( |q_\S| \leq \frac{2\pi}{\lambda_\S} \right) 
        \mathrm{rect} \left( |q_\I| \leq \frac{2\pi}{\lambda_\I} \right).
    \label{eq:q_correlation}
\end{split}
\end{equation}
This constitutes a more general form of the two-photon state than the one approximated to the paraxial regime~ \cite{Monken.1998}.
Notice that $\bra{\psi}\ket{\psi}$ corresponds to the total rate of pair generation. The diffraction limit, which states that the transverse momenta of both signal and idler is restricted by their corresponding wavelengths $|q|\leq \omega/c$ with  $\omega/c = 2\pi/\lambda$ in free-space, is modeled by the rectangular function $\mathrm{rect}(\cdot)$, being equal to one wherever its argument is true and zero otherwise. This means that modes with larger values of $|q|$ do not propagate since those correspond to evanescent modes, which do not participate in the pair-generation process, except under very special conditions \cite{Saravi.2017,santos}.
\subsection{Angular Representation}

The presence of the term $[k_{z\S}(q_\S) k_{z\I}(q_\I)]^{-1/2}$ in the joint transverse momentum amplitude in Eq.~(\ref{eq:q_correlation}) makes a physical comprehension of the photon emission more difficult, as this term diverges with $q_\S$ approaching $\omega_\S/c$ or $q_\I$ approaching $\omega_\I/c$ in the non-paraxial region.
In the following, we show analytically that such a singularity is not creating a diverging and unphysical result if we resort to an angular reference frame instead of working with the spatial frequencies $q$. As we will show, the singularity is removed in the angular representation, showing that results will be physical and converging, allowing us to perform our numerical calculations later on. Moreover, the angular representation contributes to a more natural and visual interpretation for the photon-pair generation in terms of the probability of emitting a photon pair at a particular angle for signal and idler.

\begin{figure}[t]
\centering\includegraphics{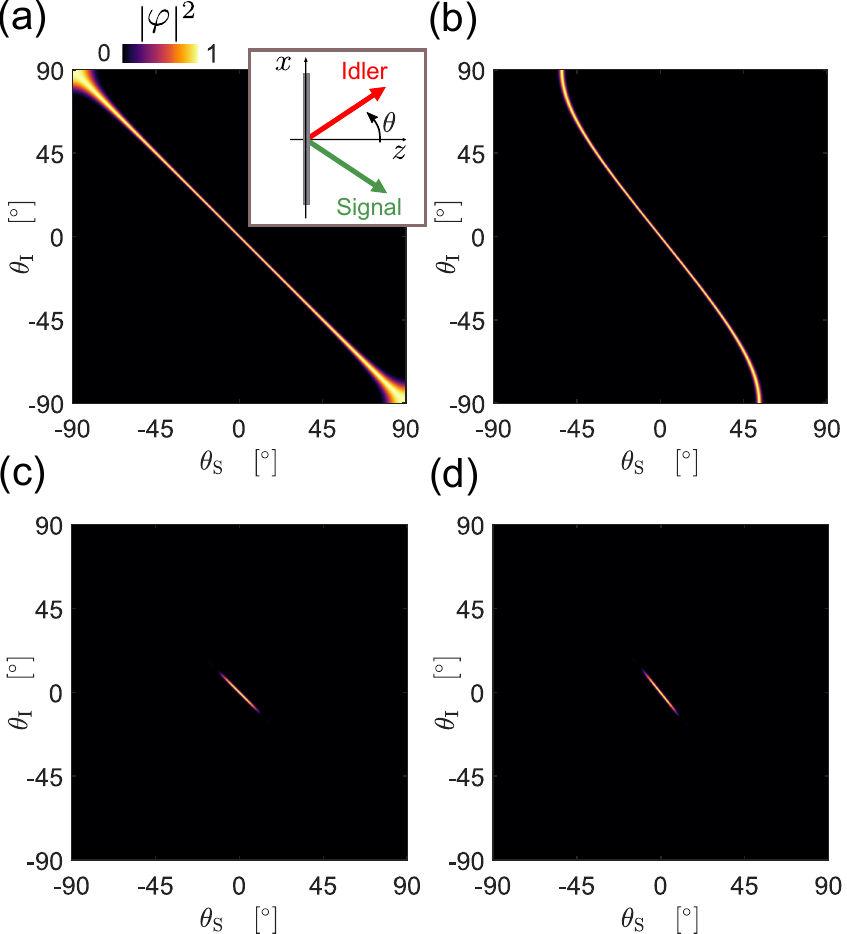}
\caption{Joint angular probability $|\varphi(\theta_\S,\theta_\I)|^2$ within $|\theta_{\S,\I}| \leq 90^{\circ}$ for signal/idler wavelengths of (a,~c)~$\lambda_\S / \lambda_\I = 1$ and (b,~d)~$\lambda_\S / \lambda_\I = 0.8$. Pump beam has a width of $10 \, \mu \mathrm{m}$. The crystal thickness $L$ is in (a,~b) much smaller than the pump, signal, and idler wavelengths ($L=3~\mathrm{nm}$), while in (c,~d) $L$ is larger than any of the three wavelengths ($L=20~\mu \mathrm{m}$). The inset shows a sketch of the angle $\theta$ that the $k$-vector makes with the $z$ axis in the $x-z$ plane.}
\label{fig:state}
\end{figure}

To go to the angular representation, consider $\theta$ as the angle between the $k$-vector and the $z$ axis in the $x-z$ plane, as shown schematically in the inset of Fig.~\ref{fig:state}(a).
With the idea of describing the ladder operators and integrals in terms of this angle instead of $q$, we start by defining $k_z=k\cos\theta$ and $q=k\sin\theta$ which gives $\dd q= \dd\theta \,k\cos\theta$ and the corresponding redefinition of the annihilation operator $\hat{a}(q,\omega,s)= (k \cos \theta)^{-1/2}\, \hat{a}(\theta,\omega,s)\,,$ which satisfies the commutation relation
\begin{equation}
    [\hat{a} (\theta,\omega,s), \hat{a}^\dagger(\theta',\omega',s')]   = \delta (\theta - \theta') \delta (\omega - \omega') \delta_{ss'} \,.
\end{equation}
Thus, we can introduce the single-frequency electric field operator [see Eq.~(\ref{eq:Eoperator})] in the angular representation as
\begin{equation}
    \begin{split}
        \hat{\mathbf{E}}^{(+)}  (x,z,\omega) = & i \left[\left( \frac{\hbar}{16 c^2 \pi^3 \varepsilon_0} \right) \left(\frac{2\pi}{l_y} \right) \right]^{1/2}
        \\ \times &
        \sum_s \int_{-\pi/2}^{\pi/2} \dd \theta \bigg[  \omega \, \hat{a}(\theta,\omega,s)
        \\ & \quad \quad  \times  
         \mathbf{e}(\theta,\omega,s) \exp\left(i qx + k_z z\right) \bigg],
        \label{eq:EoperatorAng}
    \end{split}
\end{equation}
where the diverging term is absent. The description of the electric field operator in terms of angles, following Eq.~(\ref{eq:EoperatorAng}), already includes only those modes that are non-evanescent. Notice that the non-paraxial limit is not described by large values of $q$ anymore but instead it is described by higher values of $\theta$ with angle emissions $\theta\rightarrow\pm\pi/2$.

Similarly, when introducing the angular electric field operator of Eq.~(\ref{eq:EoperatorAng}) in the two-photon state of Eq.~(\ref{eq:SPDCstate}), and calculating under the same assumptions for the polarization, we obtain the angular representation of the state
\begin{equation}
\begin{split}
    \ket{\psi} \propto \iint_{-\pi/2}^{\pi/2} \dd \theta_\S  & \dd \theta_\I 
    \; \Big[ \varphi(\theta_\S, \omega_\S; \theta_\I, \omega_\I) \;
    \\ & \times 
    \hat{a}^\dagger(\theta_\S,\omega_\S) \hat{a}^\dagger(\theta_\I,\omega_\I)\ket{0,0} \Big]\,,
    \label{eq:SPDCAng}
\end{split}
\end{equation}
where the joint angular amplitude is
\begin{equation}
\begin{split}
    \varphi & (\theta_\S, \omega_\S; \theta_\I, \omega_\I)  = \\ &
    E_{\P} \left[ \left(\frac{\omega_\S}{c}\right) \sin (\theta_\S) + \left(\frac{\omega_\I}{c}\right) \sin (\theta_I),\omega_\S+\omega_\I \right]
    \\
    & \times \mathrm{sinc}\left(\frac{\Delta k_zL}{2}\right).
    \label{eq:ang_correlation}
\end{split}
\end{equation}
Eqs.~(\ref{eq:SPDCAng}) and (\ref{eq:ang_correlation}) describe the two-photon state by using a nonlinear crystal of length $L$ that generates photon-pairs in the angular modes $\ket{\theta_\S,\theta_\I}$. It is important to emphasize that these expressions do not make any approximation on the angular range of the signal and idler photons and are also valid in the non-paraxial regime. 

The angular range of the generated pair and their correlation can be illustrated with the joint angular probability distribution $|\varphi(\theta_\S,\theta_\I)|^2$, where the degree of angular correlation is measured by the possible angular range of one photon given the emission of the other photon at one fixed angle.
Examples of $|\varphi(\theta_\S,\theta_\I)|^2$ are portrayed in Fig.~\ref{fig:state} for an ultra-thin ($L\rightarrow 0$) and a thick source, each for two cases: signal and idler with degenerate and nondegenerate wavelengths. Here, the pump beam has a Gaussian profile of transverse width $10 \, \mu \mathrm{m}$ and wavelength $\lambda_\P=500$ nm.
First, let us have a look at the case of the ultra-thin source, shown in Figs.~\ref{fig:state}(a) and (b). When $L\rightarrow 0$, the sinc-function term in Eq.~(\ref{eq:ang_correlation}) approaches 1, and the dominant term is the pump term $E_{\P} \left[q_\P=(\omega_\S/c) \sin (\theta_\S) + (\omega_\I/c) \sin (\theta_\I) \right]$. This means that in the thin-source limit, there is no limitation on the range of the generated signal and idler angles, as long as they can satisfy the transverse phase-matching condition with the plane-wave components of the pump beam. 
Figure~\ref{fig:state}(a) shows $|\varphi(\theta_\S,\theta_\I)|^2$ for the degenerate case with $\lambda_\S = \lambda_\I$, where we see that both photons can be generated in the whole range of $|\theta_{\S,\I}| \leq 90^{\circ}$. The nondegenerate case with $\lambda_\S/\lambda_\I < 1$ is shown in Fig.~\ref{fig:state}(b). Here, we see that the longer wavelength idler photons are again generated in the whole range of $|\theta_{\I}| \leq 90^{\circ}$, but the signal photons can only be generated in a limited range. This limited range is determined by the transverse phase-matching condition. Consider the simple case where the pump is almost a plane wave with $q_\P=0$, resulting in the transverse phase-matching condition $(\omega_\S/c) \sin (\theta_\S) + (\omega_\I/c) \sin (\theta_I) =0$ for the signal and idler photons, or otherwise written $\sin(\theta_\S)=-(\lambda_\S/\lambda_\I) \sin(\theta_\I)$. It is clear from this expression, that for degeneracy factors $\lambda_\S/\lambda_\I < 1$, the possible angular generation range of the shorter-wavelength signal photon will be restricted to $\theta_\S^\mathrm{max}=|\sin^{-1} [(\lambda_\S/\lambda_\I) \sin(\theta_\I=90^{\circ}) ]|= |\sin^{-1} (\lambda_\S/\lambda_\I)|$ . Intuitively, this is due to the fact that the magnitude of the wave-vector for the longer-wavelength idler photons is smaller than that of the shorter-wavelength signal, and cannot satisfy the transverse phase-matching condition with the signal photon after $\theta_\S^\mathrm{max}$, hence no signal photons will be generated after that angle.
As we will see, this physical effect will play the key role in setting the diffraction limited resolution of QIUP.

Additionally, we notice that the degree of correlation of signal and idler is higher for small angles than for very large angles.
A decreased correlation will certainly lead to a reduced resolution in a non-local imaging system \cite{Abouraddy.FourierOptics.2002}. 
For pushing the resolution to its limit and also to concentrate on the effect of the wavelength of signal and idler on the resolution, we will study the case where signal an idler are created from a plane wave pump which ensures perfect transverse correlation between them. Additionally, we later on analyze the effect of a degraded correlation on the resolution by using a pump with finite width.

Finally, we look at $|\varphi(\theta_\S,\theta_\I)|^2$ for the case of a thicker crystal ($L=20~\mu \mathrm{m}$), in the degenerate and nondegenerate cases, shown in Figs.~\ref{fig:state}(c) and (d), respectively. As can be seen, the angular generation ranges for both signal and idler photons are strongly reduced, which is caused by the longitudinal phase-matching effect embedded in the sinc-function of Eq.~(\ref{eq:ang_correlation}). As we will show in our calculations, the thin source provides the ultimate possible resolution, which is directly related to the fact that it can generate the largest possible range of transverse momentum modes for the photon pair, or equivalently the widest range of generation angles.
\section{QIUP beyond the paraxial regime: Formulation}
\label{sec:QIUP}
The core of QIUP~\cite{LEMOS.QIUP.2014, LAHIRI.TheoryQIUP.2015} is the effect of induced coherence without induced emission \cite{MANDEL.IndCoh1.1991, MANDEL.IndCoh2.1991}. A sketch of the scheme is illustrated in Fig.~\ref{fig:setup}, consisting of two photon-pair sources A and B forming a non-linear interferometer \cite{CHEKHOVA.NonlinInterf.2016}. The object interacts only with the idler photons of source~A, then the path of these photons is aligned with the path of the idler photons of source~B. This alignment introduces interference in the state of the system since now we cannot distinguish which source generated the photon-pair. The signal photons of both sources, that never interact with the object, interfere at the beam splitter and are measured by the camera revealing the object, while the idler photons remain undetected. 
In this section, we derive the expression of the photon-counting rate of this non-local image similar to the treatment of Ref.~\cite{LAHIRI.QIUPPosition.2021}; however, our analysis is not restricted to the paraxial regime and consequently can be used to evaluate the ultimate resolution of quantum imaging, which as we will show, can be achieved by using thin sources and in a strongly non-paraxial regime of operation.
It should be noted that for derivation of the analytical expression we resort to the $q$-domain representation, mainly due to the fact that Fourier transform expressions have a simpler form in this domain. For numerical evaluations and interpretation of the physics we convert to the angular-domain, which yields singularity-free expressions.


Considering Eqs.~(\ref{eq:SPDCstate_q}) and (\ref{eq:q_correlation}), the quantum state of both sources A and B is then $\ket{\psi} = \ket{\psi}_\A + \ket{\psi}_\B$. Since the following derivation takes into account the low-gain regime, at most one pair is present in the system at a time \cite{LAHIRI.SINGLEPHOTONINT.2017}. As sketched in the inset of Fig.~\ref{fig:setup}, each position $x_\I$ of the object is modeled by a lossless beam-splitter with a transmission $T(x_\I)$ and reflection $R(x_\I)$, one of the inputs being $\hat{E}^{(+)}_{\I,\A}(x_\I)$ and the other the vacuum $\hat{E}^{(+)}_0(x_\I)$. The collinear output with $\hat{E}^{(+)}_{\I,\A}(x_\I)$ that goes towards source~B is
\begin{equation}
    \hat{E}^{(+)}_{\I,\B}(x_\I) =  T(x_\I) \hat{E}^{(+)}_{\I,\A}(x_\I) + R(x_\I) \hat{E}^{(+)}_0(x_\I),
    \label{eq:T_BS}
\end{equation}
which coincides with the spatial modes of the idler of source~B given that the idler arms are aligned. Since the beam-splitter is lossless, then $T(x) \, T^*(x) + R(x) \, R^*(x)=1$. Furthermore, the electric field operator can be also simplified following the approximations mentioned in the two-photon state at the end of Sec.~\ref{sec:twophoton}, so Eq.~(\ref{eq:Eoperator}) becomes
\begin{equation}
\begin{split}
        \hat{E}^{(+)}_{\I,j}(x_\I) \propto \int & \dd q_\I \bigg[ \exp(i q_\I x_\I) (k_{z \I})^{-1/2} 
        \\ & \times
        \mathrm{rect} \left( |q_\I| \leq \frac{2\pi}{\lambda_\S} \right) \hat{a}_{j}(q_\I) \bigg]
\end{split}
    \label{eq:Eop_simplified}
\end{equation}
with $j \in \{\A,\B,0\}$. The limits of any integral, from here on, are $\pm \infty$ unless otherwise noted.  Thus, we can find the relation of the diffraction-limited idler modes of source~B with the object and the idler modes of source~A. To this end, we put Eq.~(\ref{eq:Eop_simplified}) into Eq.~(\ref{eq:T_BS}) and use the Fourier transform of the transmission $\widetilde{T}(q_\I) = \int \dd x_\I T(x_\I) \exp(i q_\I x_\I)$ and likewise for the reflection, which results in 
\begin{equation}
\begin{split}
        \hat{a}_{\B} & (q_\I) \, (k_{z\I})^{-1/2} \, \mathrm{rect} \left( |q_\I| \leq \frac{2\pi}{\lambda_\I} \right) 
        \\ & = \int \dd q'_\I  \Bigg\{ \Big[ \widetilde{T}(q_\I - q'_\I) \hat{a}_{\A}(q'_\I) + \widetilde{R}(q_\I - q'_\I) \hat{a}_0(q'_\I)\Big]
        \\ & \quad \quad \times (k'_{z\I})^{-1/2} \mathrm{rect} \left( |q'_\I| \leq \frac{2\pi}{\lambda_\I} \right) \Bigg\}.
        \label{eq:idlermodes_matched}
\end{split}
\end{equation}
Importantly, the rectangular function ensures that only propagating modes of the idler are included.

To exploit the position correlations for imaging, signal photons are measured by a camera located in the image plane of both sources, these signal photons interfere in a $50:50$ beam splitter. Then, the electric field operator at the camera is
\begin{equation}
    \begin{split}
         \hat{E}^{(+)}_\mathrm{cam} \propto \int &  
         \dd q_\S 
         \bigg\{  \exp\left(iq_\S x_\S \right) \; k_{z\S}^{-1/2}  
         \\ & \times  
         \mathrm{rect} \left( |q_\S| \leq \frac{2\pi}{\lambda_\S} \right) 
         \\ & \times 
         \left[\hat{a}_{\A}(q_\S) + i \exp(i \eta) \hat{a}_{\B}(q_\S) \right] \bigg\},
     \end{split}
     \label{eq:Eoper_camera}
\end{equation}
where $\eta$ is an accumulated phase difference between the two signal arms.
Lastly, the photon counting rate at the camera is found by $\mathcal{R}(x_\S) \propto \bra{\psi} \hat{E}^{(-)}_\mathrm{cam} \hat{E}^{(+)}_\mathrm{cam} \ket{\psi}$, resulting in
\begin{equation}
    \mathcal{R}(x_\S) \propto \int \dd x_\I \left[ \,|\Phi_\A|^2 + |\Phi_\B|^2 + 2 \mathrm{Re} \left( \Phi^*_\A \Phi_{\B \widetilde{T}} \right) \right],
    \label{eq:R_PC}
\end{equation}
with
\begin{equation}
    \begin{split}
        \Phi_\A  (x_\S, x_\I) & = \iint \dd q_\I \dd q_\S \bigg[ \exp(iq_\S x_\S + iq_\I x_\I)
        \\ & \quad \quad \quad \quad  \times
        (k_{z \S})^{-1/2} \phi_\A(q_\S,q_\I) \bigg] , 
        \\
        \Phi_\B  (x_\S, x_\I) & =  \iint \dd q_\I \dd q_\S \bigg[ \exp(iq_\S x_\S + iq_\I x_\I) 
        \\ & \quad \quad \quad \quad  \times
        (k_{z \S})^{-1/2} \phi_\B(q_\S,q_\I) \bigg], 
        \\
        \Phi_{\B \widetilde{T}} (x_\S, x_\I) & = \iint \dd q_\I \dd q_\S \bigg[ \exp(iq_\S x_\S + iq_\I x_\I) 
        \\ & \quad \quad \quad \times
        (k_{z \S} k_{z \I})^{-1/2} 
        \mathrm{conv}(q_\S,q_\I) 
        \\ & \quad \quad \quad \times
        \mathrm{rect} \left( |q_\I| \leq \frac{2\pi}{\lambda_\I} \right) \bigg],
    \end{split}
    \label{eq:R_parts}
\end{equation}
where $\mathrm{conv}(q_\S,q_\I)$ denotes the following convolution $\circledast$ along $q_\I$
\begin{equation}
    \begin{split}
        \mathrm{conv} & (q_\S,q_\I) = \Big[ (k_{z\I})^{1/2} \,\phi_\B  (q_\S,q_\I)\Big]  \circledast \widetilde{(T^*)}(q_\I)  
        \\  = & 
        \int \dd q'_\I \, (k'_{z \I})^{1/2} \, \phi_\B(q_{\S}, q'_{\I})  \, \widetilde{T}^*(q'_\I - q_\I).
    \end{split}
\end{equation}
Additionally, we take $\eta = -\pi/2$ to have constructive interference at the camera (see Appendix~\ref{app:R} for more details of the derivation).

The first two terms of Eq.~(\ref{eq:R_PC}) correspond to the individual contribution of each of the sources as if the other source was absent. Therefore, these two terms do not carry information of the object and are irrelevant for the resolution analysis, from here on we will take into account only the interference term that contains the image
\begin{equation}
       \mathcal{I}(x_\S) \propto \int \dd x_\I \,  \mathrm{Re} \left( \Phi^*_\A \Phi_{\B \widetilde{T}}  \right).
     \label{eq:image}
\end{equation}
This term corresponds to the joint response of the sources. In practice, the background $\int \d x_\I \left( |\Phi_\A |^2 + |\Phi_\B |^2 \right) $ on top of the image $\mathcal{I}$ can be removed by taking advantage of the phase difference of both output ports of the beam splitter and subtracting one image from the other \cite{LEMOS.QIUP.2014,LAHIRI.TheoryQIUP.2015}.

The resolution of a conventional imaging system depends on the illumination wavelength and also on the numerical aperture of the optical elements \cite{born_wolf}. In this work, we focus on evaluating the resolution of QIUP at its ultimate limit, therefore, we assume that the optical elements in Fig.~\ref{fig:setup} have a numerical aperture equal to one and we focus on the dependence of the transverse resolution on the signal and idler wavelengths $\lambda_{\S,\I}$. To derive the resolution, we follow Rayleigh's criterion and assume the object consists of two infinitely thin slits that are separated by a distance $d$, namely with a transmission $T(x_\I) = \delta( x_\I -d/2) + \delta(x_\I +d/2)$, and we find their image through Eq.~(\ref{eq:image}). Additionally,
we model the pump with a Gaussian profile of width $\sigma_\P$ to analyze realistic experimental scenarios where the correlation of signal and idler is imperfect, therefore, $E_\P(q_\P) = \exp (-\sigma_\P^2 q_\P^2 /2)$.
Incorporating the aforementioned criteria in conjunction with Eq.~(\ref{eq:q_correlation}), then the convolution term in $\Phi_{\B \widetilde{T}}$ becomes
\begin{equation}
    \begin{split}
         \mathrm{conv} & (q_\S,q_\I) \propto (k_{z\S})^{-1/2} \, \mathrm{rect} (|q_\S| \leq 2\pi/\lambda_\S)  
        \\   \times &
         \mathrm{Re} \Bigg\{ \exp \left( i \frac{d}{2}q_\I \right) \int \dd q'_\I \exp \left( -i \frac{d}{2}q'_\I \right)  \\ 
        & \times \mathrm{sinc}\left[ \frac{\Delta k_z(q_\S, q'_\I) L_\B}{2} \right] \exp \left[ -\frac{\sigma_\P^2}{2} (q_\S + q'_\I)^2 \right]     \\
        & \times \mathrm{rect}(|q'_\I| \leq 2\pi/\lambda_\I) \Bigg\}.
    \end{split}
    \label{eq:conv}
\end{equation}
To find the fundamental limitation of the resolution, we first analyze the case where signal and idler have the strongest possible correlation, i.e., using a plane wave pump where $\sigma_\P \to \infty$,
\begin{equation}
    \begin{split}
        \mathrm{conv} & (q_\S,q_\I) \propto 
        \\ & 
        (k_{z\S})^{-1/2} \; \cos{\left[ \frac{d}{2}\left(q_\S + q_\I \right) \right]}  \\ 
        & \times \mathrm{sinc}\left\{ \frac{L_\B}{2} \left[ \frac{2\pi}{\lambda_\P} - k_{z\S}  - \kappa  \right] \right\} \\
        & \times \mathrm{rect} \left[ |q_\S| \leq 2\pi \; \mathrm{min}\left(\frac{1}{\lambda_\S}, \frac{1}{\lambda_\I} \right) \right]  
    \end{split}
    \label{eq:q_sourceB}
\end{equation}
with $\kappa \coloneqq \left[(2\pi/\lambda_\I)^2 - q_\S^2 \right]^{1/2}$ and importantly 
\begin{equation}
    \begin{split}
        \mathrm{rect} (|q_\S| & \leq 2\pi/\lambda_\S ) \, \mathrm{rect} \left(|q_\S| \leq 2\pi/\lambda_\I \right)  \\ & = \mathrm{rect} \left[ |q_\S| \leq 2\pi \; \mathrm{min}\left(1/\lambda_\S, 1/\lambda_\I \right) \right].
        \label{eq:qs_limit}
    \end{split}
\end{equation}
The information about the distance between the slits $d$ is imprinted in the cosine term of Eq.~(\ref{eq:q_sourceB}); the closer the slits, the smaller the cosine frequency in the $q$-domain. If the cosine were to spread infinitely with all possible transverse momentum modes, infinitely close slits can be resolved in the image.
However, as it was shown in the previous section, a photon pair source can only generate modes with a restricted range of transverse momentum, where the limit is set by the thickness of the source and the wavelengths of the signal and idler photons. 
Hence, the minimum resolvable slit distance, and more generally the imaging resolution, will be limited to this available range of transverse wave-vectors.
This can also be seen in Eqs.~(\ref{eq:R_parts}) and (\ref{eq:q_sourceB}), which we derived for quantum imaging beyond the paraxial limit. 
In these equations, the hard limit for the range of available transverse wave-vectors for the detected signal photons is $2\pi \; \mathrm{min}\left(1/ \lambda_\S, 1/ \lambda_\I \right)$, which means that the limit is set by the larger of the signal and idler wavelengths. Hence, the capability in resolving the slit distance that appears in the cosine term of Eq.~(\ref{eq:q_sourceB}) will be restricted by the diffraction-limit of the larger wavelength of the signal and idler. This fact, which is the main physical finding of our work and can be intuitively seen from our derived analytical expressions, will also be verified in our following numerical calculations of the image of the two slits.

For our numerical calculations, we switch to the angular representation, to avoid the singularity terms that appear in the $q$-domain expressions derived in this section.
The functions $\Phi_\A$ and $\Phi_{\B\widetilde{T}}$ can be expressed in angular coordinates, as shown in Sec.~\ref{sec:twophoton}, where we take $q=(2\pi/\lambda) \sin(\theta)$, $k_z=(2\pi/\lambda) \cos(\theta)$ and $\dd q = k_z \dd \theta$. Thus, the angular representation of Eq.~(\ref{eq:R_parts}) becomes
\begin{equation}
    \begin{split}
        \Phi_{\A} & (x_\S,x_\I)  = 
        \iint_{-\pi/2}^{+\pi/2} \dd \theta_\S \dd \theta_\I  \Bigg\{ \varphi_{\A}(\theta_\S, \theta_\I) 
        \\ & \times \exp \left[ \frac{2\pi}{\lambda_\S} \sin(\theta_\S) x_\S + \frac{2\pi}{\lambda_\I} \sin(\theta_\I) x_\I \right] \Bigg\},
        \\
        \Phi_{\B\widetilde{T}} & (x_\S,x_\I)  =  \iint_{-\pi/2}^{+\pi/2} \dd \theta_\S \dd \theta_\I \Bigg\{ \varphi_{\B\widetilde{T}}(\theta_\S, \theta_\I)
        \\ & \times \exp \left[ \frac{2\pi}{\lambda_\S} \sin(\theta_\S) x_\S + \frac{2\pi}{\lambda_\I} \sin(\theta_\I) x_\I \right] \Bigg\},
    \end{split}
\end{equation}
where
\begin{equation}
    \begin{split}
        {\varphi}_\A & (\theta_\S, \theta_\I)  = \left[ \frac{2\pi}{\lambda_\I} \cos{(\theta_\I)} \right]^{1/2}
        E_\P (\theta_\S, \theta_\I) \,
        \Pi (\theta_\S, \theta_\I, L_\A),
        \\
        {\varphi}_{\B\widetilde{T}} & (\theta_\S, \theta_\I)  =  \left[ \frac{2\pi}{\lambda_\I} \cos{(\theta_\I)} \right]^{1/2}  
        \mathrm{conv} (\theta_\S, \theta_I)\,,
    \end{split}
    \label{eq:q_sourceB_ang}
\end{equation}
with 
\begin{equation}
\begin{split}
        E_\P & (\theta_\S, \theta_\I) = \exp \left( -\frac{\sigma_\P^2}{2} \left[ \frac{2\pi}{\lambda_\S} \sin{(\theta_\S)} + \frac{2\pi}{\lambda_\I} \sin{(\theta_\I)} \right] ^2 \right),
        \\
        \Pi & (\theta_\S, \theta_\I, L) = 
        \\ &
        \mathrm{sinc} \left\{ \frac{L}{2} \left[ k_{z \P}(\theta_\S, \theta_\I) - \frac{2\pi}{\lambda_\S} \cos{(\theta_\S)}  - \frac{2\pi}{\lambda_\I} \cos{(\theta_\I)}  \right] \right\},
        \\
        k_{z \P} & (\theta_\S, \theta_\I) =  \left\{ \left( \frac{2\pi}{\lambda_\P} \right)^2 - \left[ \frac{2\pi}{\lambda_\S} \sin(\theta_\S) + \frac{2\pi}{\lambda_\I} \sin(\theta_\I) \right]^2 \right\}^{1/2},
        \\
         \mathrm{conv} & (\theta_\S, \theta_I) \propto
         \mathrm{Re} \Bigg\{ \exp \left( i \frac{d}{2} \left[ \frac{2\pi}{\lambda_\I} \sin(\theta_\I) \right] \right) 
         \\ \times &
         \int_{-\pi/2}^{+\pi/2} \dd \theta'_\I \, \bigg[ \frac{2\pi}{\lambda_\I} \cos(\theta'_\I) 
         \exp \left( -i \frac{d}{2} \left[ \frac{2\pi}{\lambda_\I} \sin(\theta'_\I) \right] \right)
         \\  & \times
         E_\P (\theta_\S, \theta'_\I) \,
        \Pi (\theta_\S, \theta'_\I, L_\B) \bigg] \Bigg \}.
\end{split}
\end{equation}
Furthermore, to reduce the computational effort of numerically solving several integrals, an alternative expression of the image can be found by plugging $\Phi_\A$ and $\Phi_{\B\widetilde{T}}$ into the image expression of Eq.~(\ref{eq:image}). To arrive to a simplified expression, we first solve analytically the integral $\int \dd x_\I$ and then solve one of the integrals of $\int_{-\pi/2}^{+\pi/2} \dd \theta_\I$. This results in
\begin{equation}
\begin{split}
\mathcal{I} & (x_\S)  \propto \iint_{-\pi/2}^{+\pi/2} \dd \theta_\S \dd \theta_\I \, \mathrm{Re} \Bigg\{ \bigg | \frac{2\pi}{\lambda_\I} \cos{(\theta_\I)} \bigg | ^{-1}
\\ & \times {\varphi}_{\B\widetilde{T}} (\theta_\S, \theta_\I) \exp \left[i \frac{2\pi}{\lambda_\S} \sin(\theta_\S) x_\S \right] 
\\ & \times
 \int_{-\pi/2}^{+\pi/2} \dd \theta'_\S \, {\varphi}^*_\A (\theta'_\S, \theta_\I) \exp \left[ - i \frac{2\pi}{\lambda_\S} \sin(\theta'_\S) x_\S \right]  \Bigg\}.
\end{split}
\label{eq:I_reduced}
\end{equation}
As can be seen, the resulting expressions in the angular domain do not have diverging singularities [notice that the term $|(2\pi/\lambda_\I) \cos(\theta_\I)|^{-1}$ in Eq.~(\ref{eq:I_reduced}) cancels out with the two  $[(2\pi/\lambda_\I) \cos(\theta_\I)]^{1/2}$ terms in $ \varphi_\A$ and $ \varphi_{\B \widetilde{T}}$ of Eq.~(\ref{eq:q_sourceB_ang})]. These expressions are the main analytical findings of our work and will be used in the next section to numerically determine the ultimate resolution of QIUP.

\section{Numerical calculations}
\label{sec:resolution}
In this section, we numerically evaluate the ultimate resolution of QIUP, based on the non-paraxial expressions found in the previous section for a double-slit object. We first take a close look at the difference in resolution between a QIUP system with thick and ultra-thin SPDC sources in a highly nondegenerate scenario. Afterwards, we do a systematic calculation of the resolution of QIUP as a function of the thickness of the SPDC sources, showing that the ultimate resolution is reached for sources thinner than $\mathrm{max}(\lambda_\S, \lambda_\I)$, with the ultimate resolution being approximately $\mathrm{max}(\lambda_\S, \lambda_\I)/2$.  We also investigate the effect of the pump width on the resolution which is relevant for realistic experimental conditions.

\subsection{QIUP with thick and ultra-thin crystals}
\begin{figure}[t]
\centering\includegraphics{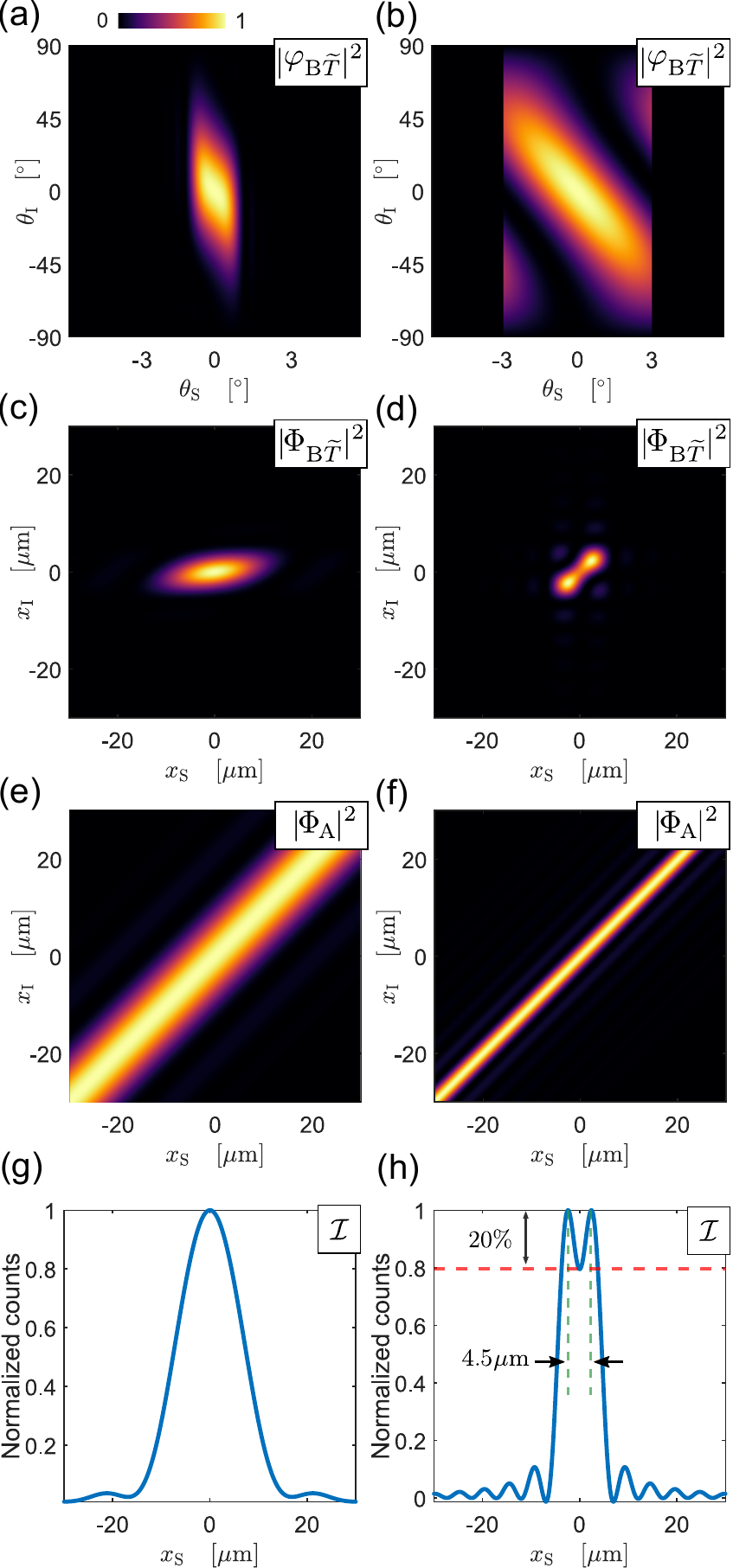}
\caption{
(a,~b)~Angular correlation $|\varphi_{\B\widetilde{T}}|^2$, (c-f)~spatial correlations $|\Phi_{\B\widetilde{T}}|^2$, $|\Phi_\A|^2$ and (g,~h)~image $\mathcal{I}$, using $\lambda_\I = 10~\mathrm{\mu m}$, $\lambda_\S =  530~\mathrm{nm}$, plane wave pump, and slits distance $d = 4.5 ~\mathrm{\mu m}$. Left column: $L = 100~\mathrm{\mu m}$ and right column: $L = 100~\mathrm{nm}$. }
\label{fig:steps}
\end{figure}

We showcase a numerical example in Fig.~\ref{fig:steps} to demonstrate the role of the crystal thickness in the resolution, here both crystals have identical thicknesses $L_\A = L_\B = L$. The object under study is composed of two slits separated by a distance $d = 4.5 ~\mathrm{\mu m}$.
We want to image this object at the wavelength of $\lambda_\I = 10~\mathrm{\mu m}$. To do this, we use a pump at the wavelength of $\lambda_\P \approx 503 ~\mathrm{nm}$, which based on the conservation of energy requires us to detect signal photons at $\lambda_\S =  530~\mathrm{nm}$ on the camera.
We illustrate two cases: (left column) a  thick source with $L = 100~\mathrm{\mu m}$, which is longer than both signal and idler wavelengths, (right column) an ultra-thin source $L = 100~\mathrm{nm}$, much shorter than the wavelengths. For the sake of clarity, we depict not only the resulting image $\mathcal{I}(x_\S)$, but also the correlation functions $|\varphi_{\B\widetilde{T}}(\theta_\S, \theta_\I)|^2$, $|\Phi_{\B\widetilde{T}}(x_\S,x_\I)|^2$, and $|\Phi_\A (x_\S,x_\I)|^2$ that lead to it.

Figure~\ref{fig:steps}(a) depicts the resulting $|\varphi_{\B\widetilde{T}}(\theta_\S, \theta_\I)|^2$ with a thick crystal. The signal and idler photons generated from this crystal, following the treatment of Sec.~\ref{sec:twophoton}, cover only a small range of angles ($\theta_\S^\mathrm{max} \approx \pm 1.5^\circ$ and $\theta_\I^\mathrm{max} \approx \pm 21.5^\circ$), mainly due to the longitudinal phase matching in the thick crystal. The idler photons in Fig.~\ref{fig:steps}(a) display, in turn, a larger angular range since $\varphi_{\B\widetilde{T}}$ includes the interaction of these idler photons with infinitely thin slits, which diffracts them in a wider angular range.
In contrast, Fig.~\ref{fig:steps}(b) shows that due to the use of an ultra-thin crystal and lack of restriction from longitudinal phase matching, the signal photons can be produced in their maximal range set by the degeneracy factor. This limited range is set by the rectangular function in Eq.~(\ref{eq:qs_limit}), resulting in $|\sin(\theta_\S)| \leq \min (1, \lambda_\S/\lambda_\I)=530~\mathrm{nm} / 10~\mathrm{\mu m}$, which gives $\theta_\S^\mathrm{max} \approx \pm 3^\circ$. The idler photons are produced within the whole  $\pm 90^\circ$ range.
This increased angular range now includes the side lobes, appearing in the corners of Fig.~\ref{fig:steps}(b), which carry information about the distance of the two slits.
The corresponding spatial correlation $|\Phi_{\B\widetilde{T}}(x_\S,x_\I)|^2$ is illustrated in Figs.~\ref{fig:steps}(c,~d), where the information that the object is composed of two slits is clear only in the case with an ultra-thin crystal in Fig.~\ref{fig:steps}(d). Additionally, $|\Phi_\A(x_\S,x_\I)|^2$ is illustrated in Figs.~\ref{fig:steps}(e,~f), which corresponds to the joint spatial probability distribution of source A.
It should be pointed out, that in the numerics, the plane-wave pump has been approximated with a Gaussian function of $1~\mathrm{m}$ width to numerically implement a very strong signal and idler correlation. Later on, we will discuss the effect of the pump width.

Lastly, Figs.~\ref{fig:steps}(g,~h) illustrate the resulting image $\mathcal{I}(x_\S)$ according to Eq.~(\ref{eq:image}) which is found by first taking the overlap of the position modes, $\mathrm{Re} \left( \Phi^*_\A \Phi_{\B \widetilde{T}}  \right)$, and then integrating over $x_\I$. This numerical example showcases that the use of the thick crystal on the left column fails to resolve such a small distance between the slits, whereas QIUP with the ultra-thin SPDC sources is able to resolve them. 

\subsection{Diffraction-limited resolution}
In this part, we perform a systematic calculation of the achievable resolution of QIUP as a function of the thickness of the SPDC sources, based on the described double-slit object and the highly non-degenerate wavelengths for the pair. 
Here, we abide to the heuristic convention that there should be at least a $20\%$ dip between the maxima of the image so that the two slits can be told apart from one another \cite{born_wolf}. In fact, as can be seen in Fig.~\ref{fig:steps}(h), the example of two slits separated by $d =4.5~\mathrm{\mu m}$ that was treated in the previous sections corresponds to this condition. Hence, based on the $20\%$ dip condition, the minimum resolvable distance of the slits for QIUP with photon-pair wavelengths of $\lambda_\I= 10~\mathrm{\mu m}$ and $\lambda_\S= 530~\mathrm{nm}$ is $d_\mathrm{min} =4.5~\mathrm{\mu m}$.
%
\begin{figure}[t]
\centering\includegraphics{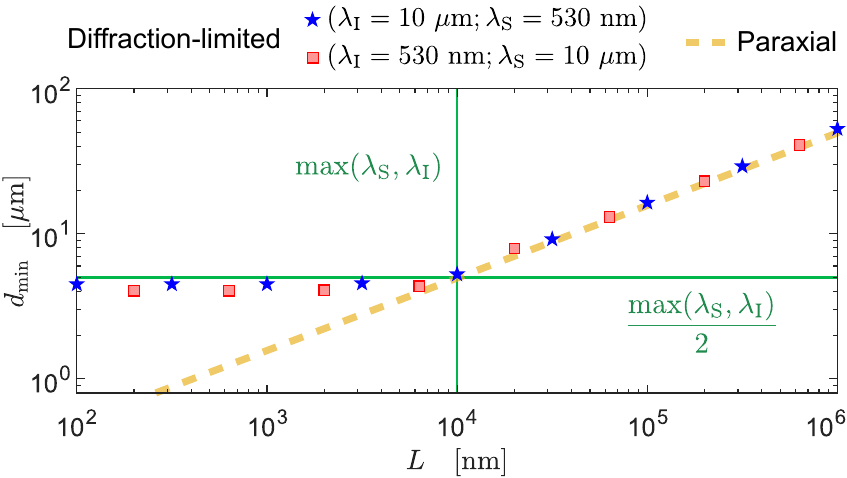}
\caption{Minimum resolvable distance $d_\mathrm{min}$ with respect to the crystal thickness $L=L_\A=L_\B$ using a plane wave pump. The diffraction-limited model is represented with blue stars and red squares, for the former, the idler wavelength is $\lambda_\I= 10~\mathrm{\mu m}$ and the signal is $\lambda_\S= 530~\mathrm{nm}$, for the latter, the wavelengths are exchanged. The paraxial estimate of $d_\mathrm{min}$ according to Eq.~(\ref{eq:dmin_parax}) is displayed with the yellow dashed line. The green lines are used for comparison, the vertical one marks $\mathrm{max}(\lambda_\S, \lambda_\I)$ and the horizontal $\mathrm{max}(\lambda_\S, \lambda_\I)/2 = 5~\mathrm{\mu m}$.}
\label{fig:equalcrystals}
\end{figure}

In Fig.~\ref{fig:equalcrystals}, we depict the numerically evaluated $d_\mathrm{min}$ for various crystal thicknesses $L=L_\A=L_\B$, where we consider identical SPDC sources. This result has been calculated numerically with our non-paraxial model presented in Sec.~\ref{sec:QIUP}.
The blue stars show the case when the object is illuminated with the longer wavelength of $\lambda_\I=10~\mathrm{\mu m}$ and the detector measures signal photons of shorter wavelength  $\lambda_\S=530~\mathrm{nm}$, just like in the example of Fig.~\ref{fig:steps}, while the red squares show the situation with the wavelengths interchanged, $\lambda_\I=530~\mathrm{nm}$ for object illumination and $\lambda_\S=10~\mathrm{\mu m}$ for detection. To build a sense of the dimensions, vertical and horizontal green lines are also included at the crystal length of $\mathrm{max}(\lambda_\S, \lambda_\I)$ and the  minimum resolvable distance of $\mathrm{max}(\lambda_\S, \lambda_\I)/2=5~\mathrm{\mu m}$, respectively. 
Importantly, we can see that the minimum resolvable distance $d_\mathrm{min}$ becomes independent of the crystal thickness when $L \lesssim \mathrm{max}(\lambda_\S, \lambda_\I)$. Hence, to reach the ultimate resolution, the crystal thickness just has to be shorter than the longer wavelength of the signal and idler photons. This means that one does not need to use an extremely thin crystal to get the best resolution, which could lower the generation efficiency unnecessarily, but can rather use a thicker crystal to improve the overall efficiency of the process but still be in the diffraction-limited region. It should be pointed out that the exact transition value for crystal thickness that results in the diffraction-limited resolution would depend on the dispersion properties of the nonlinear crystal, but we expect it to be in the same order of $\mathrm{max}(\lambda_\S, \lambda_\I)$.

\begin{figure}[bt]
\centering\includegraphics{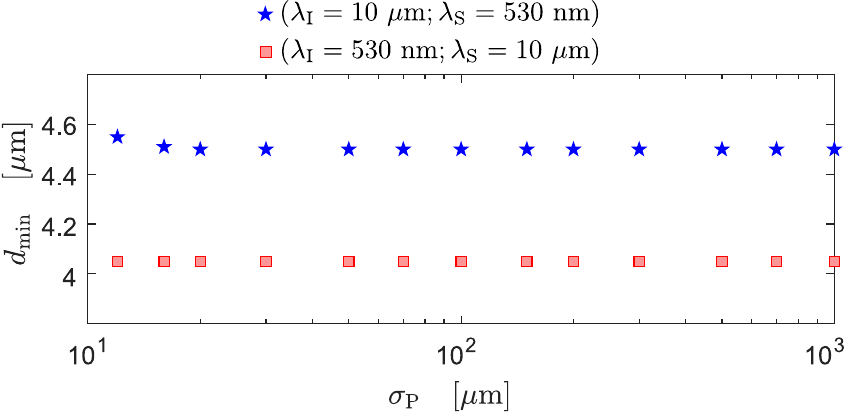}
\caption{Minimum resolvable distance $d_\mathrm{min}$ with respect to the pump width $\sigma_\P$ for a crystal thickness of $L = 100~\mathrm{nm}$.}
\label{fig:sweep_sP}
\end{figure}

To complement the results obtained with the plane-wave pump that gives the best possible correlation between signal and idler, we turn to a more realistic scenario where the pump has a finite width. We depict in Fig.~\ref{fig:sweep_sP} the dependence of the minimum resolvable distance $d_\mathrm{min}$ with the pump width $\sigma_\P$ while having an ultra-thin crystal of $L = 100~\mathrm{nm}$. Interestingly, we find that the resolution remains constant with respect to the pump width. The reason for this result is that even though a decrease in the pump width usually entails a decrease of spatial correlation between signal and idler, the use of an ultra-thin crystal ensures that the degree of spatial correlations remains high even for the small pump widths considered. This result shows that it is not necessary to have a very broad pump beam to reach the best possible resolution, but the diffraction limited resolution can also be achieved under common experimental situations.

\begin{figure}[b]
\centering\includegraphics{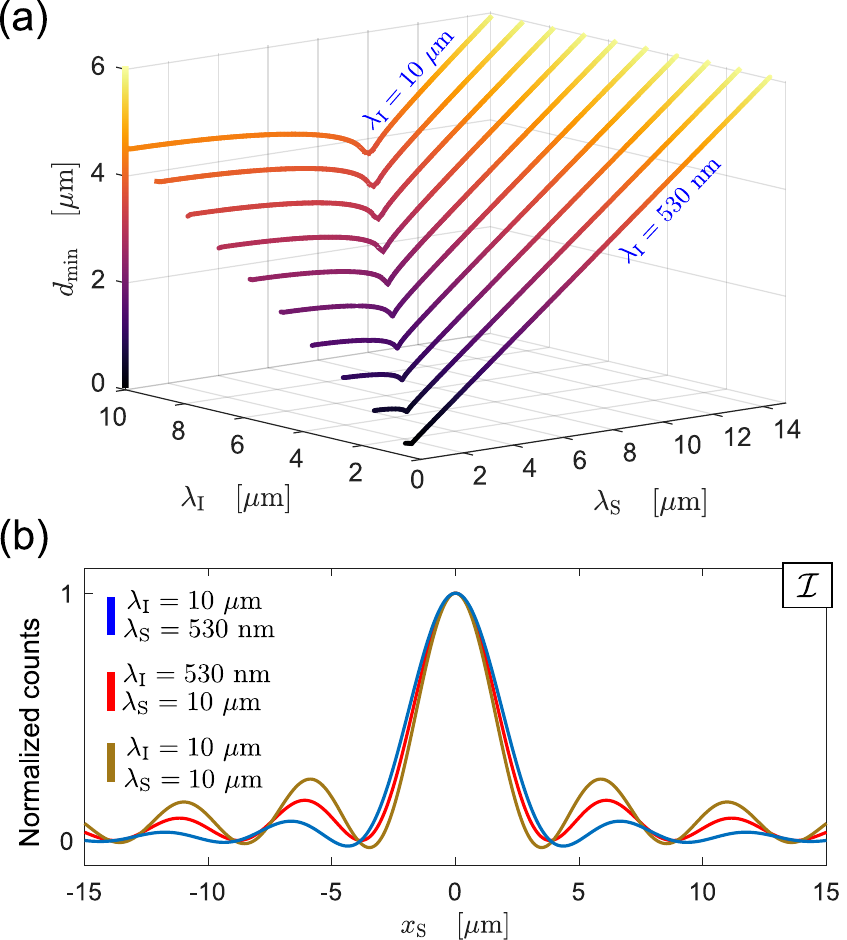}
\caption{(a) Minimum resolvable distance $d_\mathrm{min}$ with respect to the signal and idler wavelengths. Here, the pump width is $\sigma_\P= 100~\mathrm{\mu m}$ and the crystal thickness is $L = 100~\mathrm{nm}$. The minimum of $d_\mathrm{min}$ occurs in the degenerate case where $\lambda_\S = \lambda_\I$. (b) PSF of QIUP with an ultra-thin crystal of $L = 100~\mathrm{nm}$ for three different wavelength combinations.}
\label{fig:sweep_wl}
\end{figure}

Moreover, in both Fig.~\ref{fig:equalcrystals} and Fig.~\ref{fig:sweep_sP} there is clearly a slight difference in $d_\mathrm{min}$ for the two non-degenerate cases in this ultra-thin-crystal regime. To further explore this, we calculate $d_\mathrm{min}$ for various signal and idler wavelengths, maintaining a crystal thickness of $L = 100~\mathrm{nm}$ and a pump width of $\sigma_\P= 100~\mathrm{\mu m}$, see Fig.~\ref{fig:sweep_wl}(a). The idler is always the photon that illuminates the object and only the signal photon is detected by the camera. The numerical value of the minimum resolvable distance $d_\mathrm{min}$ has been assigned a color for the sake of visualization.
As expected from the previous analysis, we observe in Fig.~\ref{fig:sweep_wl}(a) that indeed the longer wavelength defines the minimum resolvable distance $d_\mathrm{min}$ and its value is close to $\mathrm{max}(\lambda_\S, \lambda_\I)/2$. A more accurate dependence of the resolution on the wavelengths can be extracted from Fig.~\ref{fig:sweep_wl}(a), namely we identify the following three prominent regimes of the diffraction-limited resolution $d_\mathrm{min}$:
\begin{equation}
     \frac{d_\mathrm{min}}{\mathrm{max}(\lambda_\S, \lambda_\I)} \approx \begin{dcases*}
                    0.45 & if $\lambda_\I \gg \lambda_\S $\\
                    0.40 & if $\lambda_\I \ll \lambda_\S $\\
                    0.37 & if $\lambda_\I = \lambda_\S $
                    \end{dcases*}.
    \label{eq:dmin_difflimit}
\end{equation}
From this analysis, one notices that the minimum of $d_\mathrm{min}$ in the degenerate case where $\lambda_\S = \lambda_\I$ shows a marginally enhanced resolution compared to the resolution in the non-degenerate scenario. The reason for the slight difference in resolution comes from the fact that the point-spread-function (PSF) is slightly different depending on the choice of wavelengths. The PSF is found by taking a single slit as the object, as depicted in Fig.~\ref{fig:sweep_wl}(b) which shows that there is a small difference in the width of the central lobe among the three PSFs. In the case of $\lambda_\I \gg \lambda_\S $, the main lobe is broader and the side lobes are small, for $\lambda_\I \ll \lambda_\S$ the central lobe gets narrower with higher side lobes, and for the degenerate case $\lambda_\I = \lambda_\S $, the main lobe is the narrowest but the side lobes are the most pronounced of the three cases. Given that the Rayleigh criterion only considers the main lobe of the image to find the minimum resolvable distance between two point-like objects, the degenerate case does result in a smaller $d_\mathrm{min}$. However, a narrower central lobe in the image $\mathcal{I}(x_\S)$ comes at the expense of having undesirable taller side lobes. Such an effect of enhanced resolution has also been observed in classical optics \cite{Rogers.2018,Chen.2019}. 

Thus, the conclusion of our analysis is twofold. First, the resulting image of the QIUP scheme with ultra-thin crystals is limited in the $q$-space by the diffraction of the longer wavelength, namely $|q_\S| \leq 2\pi \; \mathrm{min}\left(1/\lambda_\S, 1/\lambda_\I \right)$  as was shown in Eq.~(\ref{eq:qs_limit}). Second, this fundamental limit in the $q$-space leads to a resolution of $d_\mathrm{min} \approx \mathrm{max}(\lambda_\S, \lambda_\I)/2$, where the exact value depends on the wavelength combination as shown in Eq.~(\ref{eq:dmin_difflimit}). These constitute the main results of this work.

\subsection{Paraxial regime}
For the sake of completeness, we analyze the common experimental scenario where the crystal thickness is larger than the longer wavelength. Numerous works have analyzed such scenario of SPDC, e.g., Refs.~\cite{Monken.1998, SHIH.QGhostDiff.1995}, and constitutes a simplified case of our more general model. The minimum resolvable distance $d_\mathrm{min}$, as mentioned, was found numerically since analytical expressions for the position correlations $\Phi_{\B \widetilde{T}}$ and $\Phi_\A$ cannot be easily calculated using the current model. However, an analytical expression can be obtained if we assume that all the transverse momenta are small. This paraxial regime can be achieved with thick crystals where the produced signal and idler photons are generated only within a small range of angles. In such paraxial approximation, Eqs.~(\ref{eq:q_correlation}) and (\ref{eq:q_sourceB}) are greatly simplified, namely the main lobe of the sinc-function is approximated to a Gaussian, the terms $\kappa^{-1/2}$ and $k_{z}^{-1/2}$ become independent of the transverse momentum $q$ and the rectangular functions take an infinite width. 
The minimum resolvable distance $d_\mathrm{min}^\mathrm{(paraxial)}$ is then
\begin{equation}
    d_\mathrm{min}^\mathrm{(paraxial)} \approx 0.7  \left( \lambda_\S + \lambda_\I \right)^{1/2} \left( \frac{1}{L_\A} +  \frac{1}{L_\B}\right)^{-1/2},
    \label{eq:dmin_parax}
\end{equation}
see Appendix~\ref{app:paraxial} for the derivation. Equation~(\ref{eq:dmin_parax}) depends on the crystals' thicknesses and both signal and idler wavelengths and the closed-form of $d_\mathrm{min}^\mathrm{(paraxial)}$ gives physical intuition of the resolution in the paraxial regime. For the case of identical sources with $L=L_\A=L_\B$, as found in Refs.~\cite{RAMELOW.QIUPPossExp.2021,LAHIRI.QIUPosRes.2021}, then $d_\mathrm{min} \propto \left[ L(\lambda_\S + \lambda_\I) \right]^{1/2}$, showing not only that thinner crystals can improve the resolution, but also that the resolution does not depend on whether the illuminating wavelength $\lambda_\I$ is larger than the detected wavelength $\lambda_\S$, or vice versa, as shown by the sum of the two wavelengths. 
However, the paraxial result in Eq.~(\ref{eq:dmin_parax}) does not correctly describe the achievable resolution with ultra-thin crystals as it is shown in Fig.~\ref{fig:equalcrystals} where the paraxial prediction of resolution is displayed with a yellow dashed line. Essentially, the naive use of the paraxial expression for the regime with thin crystals would lead to the erroneous conclusion of having ``super-resolution''. Based on these results, we can see that the paraxial expression is only valid in the regime with thick crystals where $L \gtrsim \mathrm{max}(\lambda_\S, \lambda_\I)$.

Lastly, we find that distinguishable sources can affect the quality of the image, in particular when the sources consist of crystals that have different thicknesses. The term $\left( 1/L_\A +  1/L_\B\right)^{-1/2}$ of Eq.~(\ref{eq:dmin_parax}) suggests that making the sources more distinguishable has the consequence of reducing the resolution. This effect can be attributed to the imperfect overlap of the position modes due to the distinguishability of the sources.

\section{Discussion and summary}
\label{sec:disc}
The diffraction-limited resolution is a direct consequence of the limited range of transverse momenta set by free-space propagation, which was carefully modeled throughout this work with a formalism rigorously developed for treating the non-paraxial regime of operation.
We demonstrated that the range of transverse momenta for both signal and idler photons produced by an ultra-thin crystal is set by the longer wavelength from the pair, which consequently 
sets a limit on the achievable transverse resolution of QIUP. Furthermore, we showed that photon-pair sources with a thickness comparable or smaller than either of the signal and idler wavelengths allow to reach the diffraction-limited resolution shown in Eq.~(\ref{eq:dmin_difflimit}).
Essentially, a lack of restriction from the phase-matching condition in a thin source allows for generating the widest possible range of transverse wave-vectors, which consequently pushes the resolution to its diffraction limit.
Photon-pair generation in such ultra-thin sources has been demonstrated recently in ultra-thin nonlinear films \cite{CHEKHOVA.NonlinearFilms.2021,CHEKHOVA.WithoutMomCons.2019} and also in thin metasurfaces \cite{CHEKHOVA.SPDCMetasurface.2021}. We recognize that it is possible to achieve resolution in QIUP that goes beyond the diffraction limit if the imaging process somehow involves the participation of evanescent waves, such as the scenario proposed in Ref. \cite{santos}, where the near-field interaction of an absorptive particle at the idler wavelength is used to disturb the field at its paired signal wavelength, which allows to form a subdiffraction image with undetected photons. Yet, QIUP involving only far-field interactions, as treated in this work, is limited by the diffraction limit of the longer wavelength photon, as we have shown in our work.

It should be mentioned, that although we have performed our analysis for a narrowband signal frequency, which corresponds to the use of a narrow bandpass filter in the signal detection arm, the analysis can be easily extended to a wideband signal detection. For the case of a narrowband/continuous wave pump and thin SPDC sources, signal and idler pairs can be generated at a wide spectral range due to a lack of the longitudinal phase-matching condition, as long as they satisfy the conservation of energy. In QIUP, where we only detect the signal photons, the signal photons at one frequency do not share any phase relation to signal photons at another frequency in the SPDC generation spectrum. Hence, signal photon intensities corresponding to different frequencies add up incoherently at the detector. Hence, we can use our model to find the intensity images at each signal frequency and simply add them together. This also means that in a wideband detection of the signal, the resolution will be restricted to the longer corresponding idler wavelength.
%

Beyond QIUP, quantum ghost imaging exhibits also angular- and position correlation functions similar to the ones shown in Fig.~\ref{fig:steps}. The main difference between these schemes is that ghost imaging consists of one source of photon-pairs and measures both signal and idler in coincidences to retrieve the image \cite{Moreau.ReviewGhostImaging.2018}, $\mathcal{I} \propto \bra{\psi} \hat{E}^{(-)}_\I \hat{E}^{(-)}_\S \hat{E}^{(+)}_\S \hat{E}^{(+)}_\I \ket{\psi}$. Hence, its diffraction-limited resolution is also limited by the larger wavelength. This conclusion disagrees with Ref.~\cite{SHIH.GhostAtomicRes.2017}, that claims that quantum ghost imaging can have a resolution much smaller than the illuminating wavelength owed only to a very large non-degeneracy of photons. Finally, since the analog classical schemes of ghost imaging \cite{RoleEntangQuantClass,GhostImagClass4,ClassConjMirror,GhostImagClass3,GhostImagClass2,GhostImagClassic1} and imaging with undetected light \cite{Cardoso.ClassicalImagUndetecLight.2018} are also restricted by free-space propagation, it can be concluded that their diffraction-limited resolution will also be limited by the longer wavelength.

In summary, we constructed a theoretical formalism of photon-pair generation beyond the paraxial regime and applied it to find the minimum resolvable distance $d_\mathrm{min}$ of two infinitesimal slits in QIUP. On one hand, in the paraxial regime and for crystals' thicknesses $L$ larger than $\approx \mathrm{max}(\lambda_\S, \lambda_\I)$, we found analytically that $d_\mathrm{min}^\mathrm{(paraxial)} \propto (\lambda_\S + \lambda_\I)^{1/2}(1/L_\A + 1/L_\B)^{-1/2}$, which shows that the resolution can be improved by using thinner crystals. On the other hand, crystals thinner than $\approx \mathrm{max}(\lambda_\S, \lambda_\I)$ enable achieving the diffraction-limited resolution, which was found numerically to be $d_\mathrm{min}^\mathrm{(limit)} \approx \mathrm{max}(\lambda_\S, \lambda_\I)/2$, independent of $L$. Hence, the longer wavelength defines the maximum achievable resolution. Finally, we infer that the resolution of other schemes (e.g.,
quantum ghost imaging and their classical analogs) is also limited by the longer wavelength.

\section*{Acknowledgments}
This work was supported by the Thuringian Ministry for Economy, Science, and Digital Society and the European Social Funds (2021 FGI 0043); European Union’s Horizon 2020 research and innovation programme (Grant Agreement No. 899580); the German Federal Ministry of Education and Research (FKZ 13N14877); and the Cluster of Excellence ‘‘Balance of the Microverse’’ (EXC 2051 – project 390713860).

A. V. and E. A. S. contributed equally to this work.

\appendix
\begin{widetext}
\section{Photon counting rate at the camera}
\label{app:R}
The quantum state of two sources of photon pairs A and B is
\begin{equation}
    \ket{\psi} \propto \iint \dd q_{\S,\A}  \dd q_{\I,\A} \, \phi_\A(q_{\S,\A}; q_{\I,\A}) \hat{a}^\dagger_\A(q_{\S,\A}) \hat{a}^\dagger_\A(q_{\I,\A})\ket{0,0} + \iint \dd q_{\S,\B}  \dd q_{\I,\B} \, \phi_\B(q_{\S,\B}; q_{\I,\B}) \hat{a}^\dagger_\B(q_{\S,\B}) \hat{a}^\dagger_\B(q_{\I,\B})\ket{0,0}.
\end{equation}
Considering that the path of the idler photons are aligned, see Eq.~(\ref{eq:idlermodes_matched}), and signal photons of both sources interfere $q_{\S,\A}=q_{\S,\B}=q_\S$, then the state of the system can be written as
\begin{equation}
\begin{split}
    \ket{\psi} & \propto \iint \dd q_{\S}  \dd q_{\I} \, \phi_\A(q_{\S}; q_{\I}) \hat{a}^\dagger_\A (q_{\S}) \hat{a}^\dagger_\A(q_{\I})\ket{0,0}  
    \\ & +
    \iint \dd q_{\S}  \dd q_{\I} \, \phi_\B(q_{\S}; q_{\I}) \, (k_{z \I})^{1/2}
    \left[ \int \dd q'_\I  \widetilde{T}^*(q_\I - q'_\I) \, (k'_{z\I})^{-1/2} \mathrm{rect} \left( |q'_\I| \leq \frac{2\pi}{\lambda_\I} \right) \hat{a}^\dagger_{\A}(q'_\I) \right] 
    \hat{a}^\dagger_\B(q_{\S})\ket{0,0}
    \\ & +
    \iint \dd q_{\S}  \dd q_{\I} \, \phi_\B(q_{\S}; q_{\I}) \, (k_{z \I})^{1/2}
    \left[ \int \dd q'_\I  \widetilde{R}^*(q_\I - q'_\I) \, (k'_{z\I})^{-1/2} \mathrm{rect} \left( |q'_\I| \leq \frac{2\pi}{\lambda_\I} \right) \hat{a}^\dagger_0(q'_\I) \right]
    \hat{a}^\dagger_\B(q_{\S})\ket{0,0}.
\end{split}
\end{equation}
By rearranging the integrals of $q_\I$ and $q'_\I$ in the second and third terms,
\begin{equation}
\begin{split}
    \ket{\psi} & \propto \iint \dd q_{\S}  \dd q_{\I} \, \phi_\A(q_{\S}; q_{\I}) \hat{a}^\dagger_\A (q_{\S}) \hat{a}^\dagger_\A(q_{\I})\ket{0,0}  
    \\ & +
    \iint \dd q_{\S}  \dd q'_{\I}   \, (k'_{z\I})^{-1/2}
    \left[ \int \dd q_\I \, (k_{z \I})^{1/2} \, \phi_\B(q_{\S}; q_{\I})  \, \widetilde{T}^*(q_\I - q'_\I) \right]  \mathrm{rect} \left( |q'_\I| \leq \frac{2\pi}{\lambda_\I} \right) 
    \hat{a}^\dagger_{\A}(q'_\I) \hat{a}^\dagger_\B(q_{\S})\ket{0,0}
    \\ & +
    \iint \dd q_{\S}  \dd q'_{\I}   \, (k'_{z\I})^{-1/2}
    \left[ \int \dd q_\I \, (k_{z \I})^{1/2} \, \phi_\B(q_{\S}; q_{\I})  \, \widetilde{R}^*(q_\I - q'_\I) \right]  \mathrm{rect} \left( |q'_\I| \leq \frac{2\pi}{\lambda_\I} \right) 
    \hat{a}^\dagger_{0}(q'_\I) \hat{a}^\dagger_\B(q_{\S})\ket{0,0},
\end{split}
\label{eq:app_state}
\end{equation}
the expressions in square brackets become convolutions along $q'_\I$, denoted by $\circledast$, and taking into account that $ \widetilde{(T^*)}(q'_\I) = \widetilde{T}^*(- q'_\I)$,
\begin{equation}
    \left[(k'_{z\I})^{1/2} \,\phi_\B  (q_\S,q'_\I)\right] \circledast \widetilde{(T^*)}(q'_\I) =  \int \dd q_\I \, (k_{z \I})^{1/2} \, \phi_\B(q_{\S}; q_{\I})  \, \widetilde{T}^*(q_\I - q'_\I).
\end{equation}
Similarly, for the term that includes the reflection. 

Finally, the photon counting rate $\mathcal{R}(x_\S) \propto \bra{\psi} \hat{E}^{(-)}_\mathrm{cam} \hat{E}^{(+)}_\mathrm{cam} \ket{\psi}$ is found using the state of Eq.~(\ref{eq:app_state}) and the electric field operator at the camera, see Eq.~(\ref{eq:Eoper_camera}). To reach the expression of $\mathcal{R}$ in Eqs.~(\ref{eq:R_PC}) and (\ref{eq:R_parts}), it is important to consider that 
\begin{equation}
\bra{0,0} \hat{a}_p(q'_\I) \hat{a}^\dagger_l(q_\I) \ket{0,0} = \begin{cases} \delta(q' - q) \propto \int \dd x_\I \exp\left[ - i(q' - q) x_\I \right], & \mbox{if } p=l \\ 0, & \mbox{if } p \neq l \end{cases}
\end{equation}
\end{widetext}
with $p,l \in \{\A,0 \}$. Additionally, the relation of the transmission and reflection of a lossless beam splitter $T(x) \, T^*(x) + R(x) \, R^*(x)=1$ expressed in $q$-domain is $\widetilde{T}(q) \widetilde{T^*}(q') +\widetilde{R}(q) \widetilde{R^*}(q') = \delta(q) \delta(q')$. Thus,
\begin{equation}
\begin{split}
        \mathcal{R}(x_\S) \propto  & \int \dd x_\I \Big\{ \,|\Phi_\A|^2 + |\Phi_\B|^2 
        \\ &
        + 2 \mathrm{Re} \left[ i \exp(i \eta) \Phi^*_\A \Phi_{\B \widetilde{T}} \right] \Big\}.
\end{split}
\end{equation}
Out of convenience, the phase difference between the signals in the third term of $\mathcal{R}$ can be taken as $ \eta = -\pi/2$ to have constructive interference in the arm of the camera, while the second output of the 50:50 beam splitter would then lead to destructive interference. The terms $\Phi_\A, \Phi_\B$ and $\Phi_{\B \widetilde{T}}$ are explicitly written out in Eq.~(\ref{eq:R_parts}).

\section{Minimum resolvable distance in the paraxial regime}
\label{app:paraxial}
Thick crystals produce signal and idler photons with small transverse momenta. Therefore, in this so-called paraxial regime, Eqs.~(\ref{eq:q_correlation}) and (\ref{eq:q_sourceB}) can be significantly simplified to find an analytical expression for the minimum resolvable distance. The approximations are the following: the main lobe of the sinc-function is approximated to a Gaussian~ \cite{VEGA.PinholeGhost.2020,SCHNEELOCH.TutorialSPDC.2016, CHAN.EntanglementMigration.2007}, the terms $\kappa^{-1/2}$ and $k_{z}^{-1/2}$ become independent of the transverse momentum $q$~\cite{TSANG.QuantLitho.2007}, and the rectangular functions take an infinite width. Therefore, the convolution term in $\Phi_{\B \widetilde{T}}$ becomes
\begin{equation}
    \begin{split}
    \big[(k_{z\I} &)^{1/2}  \,\phi_\B  (q_\S,q_\I)\big] \circledast \widetilde{(T^*)}(q_\I) \\ 
    \propto & \cos \left[ \frac{d}{2} (q_\S + q_\I) \right] \exp\left[ -\gamma L_\B (\lambda_\S + \lambda_\I) \frac{q_\S^2}{8 \pi} \right],
    \end{split}
\end{equation}
where $\gamma=0.8$ ensures that the main lobe of the sinc and the Gaussian coincide at $0.1$. Furthermore, the position correlations of Eq.~(\ref{eq:R_parts}) simplify to
\begin{equation}
    \begin{split}
        \Phi_{\B \widetilde{T}} & \propto \exp \left[- 2\pi \frac{\left(x_\S + \frac{d}{2} \right)^2}{\gamma L_\B (\lambda_\S + \lambda_\I)} \right] \delta \left(x_\I + \frac{d}{2}\right)
        \\& \quad + \exp \left[- 2\pi \frac{\left(x_\S - \frac{d}{2} \right)^2}{\gamma L_\B (\lambda_\S + \lambda_\I)} \right] \delta \left(x_\I - \frac{d}{2}\right),
        \\
        \Phi_\A & \propto \exp \left[- 2\pi \frac{(x_\S - x_\I)^2}{\gamma L_\A (\lambda_\S + \lambda_\I)} \right].
    \end{split}
\end{equation}
The minimum resolvable distance $d_\mathrm{min}^\mathrm{(paraxial)}$ is found from the image $\mathcal{I}$ following Eq.~(\ref{eq:image}), when $\mathcal{I}=0.8$ at $x_\S = 0$, i.e., $20\%$ dip between the maxima, leading to Eq.~(\ref{eq:dmin_parax})
\begin{equation}
        d_\mathrm{min}^\mathrm{(paraxial)} \approx  2  \left[-\frac{\ln(0.4) \gamma}{2\pi} (\lambda_\S + \lambda_\I) \right]^{1/2} 
        \left( \frac{1}{L_\A} +  \frac{1}{L_\B}\right)^{-1/2}.
\end{equation}

\bibliography{apssamp}

\end{document}